\documentclass[aps,graphicx,10pt,twocolumn]{revtex4}
\usepackage{amssymb}
\usepackage{amsmath}
\usepackage{amscd}
\usepackage{subfigure}
\usepackage{graphicx}

\begin{document}

\title[Short Title]{Shortcuts to adiabatic passage for population transfer and maximum
entanglement creation between two atoms in a cavity}

\author{Mei Lu$^{1}$}
\author{Yan Xia$^{1,}$\footnote{E-mail: xia-208@163.com}}
\author{Li-Tuo Shen$^{1}$}
\author{Jie Song$^{2,}$\footnote{E-mail: jsong@hit.edu.cn}}
\author{Nguyen Ba An$^{3,}$\footnote{E-mail: nban@iop.vast.ac.vn}}

\affiliation{$^{1}$Department of Physics, Fuzhou University, Fuzhou
50002, China\\$^{2}$Department of Physics, Harbin Institute of
Technology, Harbin 150001, China\\$^{3}$Center for Theoretical
Physics, Institute of Physics, 10 Dao Tan, Hanoi, Vietnam}

\begin{abstract}
We use the approach of ``transitionless quantum driving'' proposed
by Berry to construct shortcuts to the population transfer and the
creation of maximal entanglement between two $\Lambda $-type atoms
based on the cavity quantum electronic dynamics (CQED) system. An
effective Hamiltonian is designed by resorting to an auxiliary
excited level, a classical driving field and an extra cavity field
mode to supplement or substitute the original reference Hamiltonian,
and steer the system evolution along its instantaneous eigenstates
in an arbitrarily short time, speeding up the rate of population
transfer and creation of maximal entanglement between the two atoms
inside a cavity. Numerical simulation demonstrates that our
shortcuts' performance is robust against the decoherences caused by
atomic spontaneous emission and cavity photon leakage.
\end{abstract}

\maketitle

\section{INTRODUCTION}

Controlling dynamics of quantum systems is of crucial importance for many
practical purposes. A widely used method is to drive the system with
external time-dependent interactions, making the total Hamiltonian, $%
H_{0}(t),$ depend explicitly on time. The change in time is managed to be
slow to allow adiabatic passage from an initial state to a target state.
Under the adiabatic following condition, it is likely that the instantaneous
eigenstates of $H_{0}(t)$ are the moving states. That is, each of them
evolves along itself all the time without transition to other ones \cite
{JPA-42-365303-2009}.

However, that is not precise. In fact, transitions between different
time-dependent instantaneous eigenstates may still happen with nonzero
probabilities which become nonnegligible if the controlling parameters do
not change slowly enough, reducing fidelity of the evolved state with
respect to the target one. Thus, controlling quantum states based on
adiabatic passage is by its nature a long process \cite{PRA-40-6741-1989}.
If the required evolution time is too long, the method may be useless,
because decoherence, noise or losses would spoil the intended dynamics. This
limits application ranges in practice, especially in the field of quantum
computing and quantum information processing where speed is of primary
concern. Therefore, accelerating the dynamics towards the perfect final
outcome is a nice idea and perhaps the most reasonable way to actually fight
against the decoherence, noise or losses that are accumulated during a long
operation time.

So far various schemes for shortcuts to slow adiabatic passage method for
arriving at a target state from an initial state have been proposed in
theory \cite{PRA-86-033405-2012,PRL-104-063002-2010,PRA-82-063422-2012,
EPL-96-60015-2011,PRL-107-016402-2011,PRA-83-013415-2011,PRA-84-043415-2011,
NJP-14-013031-2012,PRA-83-043804-2011,PRA-84-031606R-2011,SR-2-648-2012,
PRL-105-123003-2010,PRA-83-062116-2011,JPB-43-085509-2010,
NJP-14-093040-2012,arXiv-1306.0410-2013,arXiv-1309.3020-2013,arXiv-1305.5458-2013}
and implemented in experiment \cite
{PRA-82-033430-2010,EPL-93-23001-2011,PRL-109-080501-2012,
PRL-109-080502-2012,OL-37-5118-2012,Nature-8-147-2012}. To be useful, such
nonadiabatic shortcuts must, of course, be reliable, fast and robust.

Notably, as pointed out by Berry \cite{JPA-42-365303-2009}, a nearly
Hamiltonian $H(t),$ which is associated with any given reference Hamiltonian
$H_{0}(t),$ exists that derives instantaneous eigenstates of $H_{0}(t)$
exactly, i.e., transitions between them do not occur at all during the whole
duration of system evolution regardless of the rate of changing. In other
words, the instantaneous eigenstates of $H_{0}(t)$ can be regarded as truly
moving eigenstates of $H(t).$ Because of such feature of the driving Berry
called it ``transitionless quantum driving'' and $H(t)$ the
``counter-diabatic driving'' (CDD) Hamiltonian. More importantly, Berry also
worked out a general ``transitionless tracking algorithm'' to reverse
engineer $H(t)$ from $H_{0}(t).$ Recently, transitionless quantum drivings
in Berry's spirit have been experimentally demonstrated in the effective
two-level system \cite{Nature-8-147-2012}. Furthermore, Chen \emph{et al.}
\cite{PRL-105-123003-2010} have also put forward another reverse engineering
approach using the Lewis-Riesenfeld (LR) invariant to carry the eigenstates
of a Hamiltonian from a specified initial to a final configuration, then to
design the transient Hamiltonian from the LR invariant. Although different
in form, those driving methods are shown to be essentially equivalent to
each other by properly adjusting the reference Hamiltonian \cite
{PRA-83-062116-2011}.

It is worth noticing that, although the reverse engineering approach has
been applied to achieve accelerated population transfer between two internal
states of a single atom in different systems, fast population transfer
between two atoms inside a common environment has not been studied
adequately, to our knowledge. In view of the requirements for scalable
quantum computing and quantum information processing, it is desirable to
extend the approach to multi-qubit systems. In this context, the LR
invariant approach for ultrafast quantum state transfer between two $\Lambda
$-type atoms based on CQED system has just been studied in \cite
{arXiv-1305.5458-2013}.

In this paper, we present an alternative nonadiabatic proposal to speed up
the population transfer and the creation of maximal entanglement between two
atoms inside a cavity in the spirit of Berry's transitionless quantum
driving approach. Different from the previous schemes \cite
{PRL-105-123003-2010,PRA-83-062116-2011,JPB-43-085509-2010,
NJP-14-093040-2012,arXiv-1306.0410-2013} where the CDD Hamiltonian derived
from the original reference Hamiltonian can be realized experimentally in
terms of a time-dependent magnetic field between two levels of a single
atom, in our system with two atoms in a common cavity the reverse engineered
CDD Hamiltonian does not readily available within the model under
consideration. To circumvent this, we take into account one more auxiliary
excited level for each atom, an extra external laser field and an extra
cavity field mode to experimentally realize our shortcut schemes to
adiabatic passage and construct the auxiliary interaction Hamiltonian that
provides us with the extra shortcut interaction. Though the extra shortcut
interaction derived from the reference Hamiltonian can speed up the
population transfer and the rate of maximally entangled state creation, the
perfect shortcut performance is slightly deteriorated since the derivation
of the auxiliary effective Hamiltonian is based on the large detuning
condition. Still, our extra shortcut interaction can considerably accelerate
the slow adiabatic passage and the performance based on it is robust, thus
promising to be realized experimentally.

Our paper is structured as follows. In Sec. II, we describe the theoretical
model for two $\Lambda $-type atoms embedded in a single-mode cavity. In
Sec. III, we construct a shortcut passage for population transfer between
two atoms. In Sec. IV, the shortcut for generating maximal entanglement
between two atoms is achieved. In Sec. IV, the influences of decoherence on
the shortcut for population transfer and maximal entanglement generation are
considered. The conclusion part appears in Sec. VI.

\section{The model}

\begin{figure}[tbp]
\includegraphics[width=0.6\columnwidth]{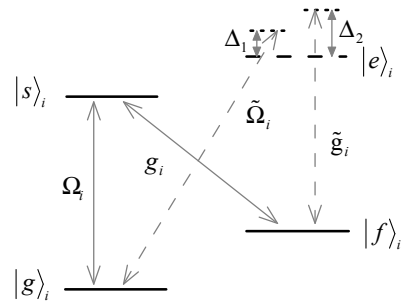}
\caption{The $\Lambda $-type configuration for the $i$th atom consists of
one excited state $|s\rangle _{i}$ and two ground states $|g\rangle _{i}$
and $|f\rangle _{i}.$ The transition $|g\rangle _{i}\leftrightarrow
|s\rangle _{i}$ is resonantly driven by a time-dependent classical field
with Rabi frequency $\Omega _{i}(t),$ while $|f\rangle _{i}\leftrightarrow
|s\rangle _{i}$ is resonantly coupled to a cavity mode $a$ with a coupling
constant $g_{i}.$ The auxiliary excited state $|e\rangle _{i}$ is
nonresonantly coupled to $|g\rangle _{i}$ by a time-dependent classical
field with Rabi frequency $\tilde{\Omega}_{i}(t)$ and to $|f\rangle _{i}$ by
a cavity mode $b$ with a coupling constant $\tilde{g}_{i}.$ $\Delta _{1,2}$\
are finite detunings.}
\label{Fig.1}
\end{figure}

The model we consider consists of two $\Lambda $-type atoms embedded in a
single-mode cavity, as sketched by solid bars and solid arrows in Fig. 1.
The transitions $|g\rangle _{1}\leftrightarrow |s\rangle _{1}$ and $%
|g\rangle _{2}\leftrightarrow |s\rangle _{2}$ are resonantly driven by two
time-dependent classical fields $\Omega _{1}(t)$ and $\Omega _{2}(t),$ which
are $3\pi /2$-dephased from each other, while each transition $|f\rangle
_{i}\leftrightarrow |s\rangle _{i}$ $(i=1,2)$ is resonantly coupled to a
cavity mode with a coupling constant $g_{i}.$ Under the rotating-wave
approximation (RWA), the time-dependent interaction Hamiltonian for the
whole system reads ($\hbar =1$):
\begin{equation}
H_{0}(t)=\Omega _{1}(t)|s\rangle _{1}\langle g|-i\Omega _{2}(t)|s\rangle
_{2}\langle g|+\sum_{i=1,2}g_{i}a|s\rangle _{i}\langle f|+H.c.,  \label{e1}
\end{equation}
where $a$ is the annihilation operator of the cavity mode.

If the initial state is $|g\rangle _{1}|f\rangle _{2}|0\rangle _{c},$ with $%
|0\rangle _{c}$ being the vacuum state of the cavity, the whole system
evolves within a single-excitation subspace spanned by five basic states
\begin{equation}
|\phi _{1}\rangle =|g\rangle _{1}|f\rangle _{2}|0\rangle _{c},  \label{e2}
\end{equation}
\begin{equation}
|\phi _{2}\rangle =|s\rangle _{1}|f\rangle _{2}|0\rangle _{c},
\end{equation}
\begin{equation}
|\phi _{3}\rangle =|f\rangle _{1}|f\rangle _{2}|1\rangle _{c},
\end{equation}
\begin{equation}
|\phi _{4}\rangle =|f\rangle _{1}|s\rangle _{2}|0\rangle _{c}
\end{equation}
and
\begin{equation}
|\phi _{5}\rangle =|f\rangle _{1}|g\rangle _{2}|0\rangle _{c}.  \label{e6}
\end{equation}

We assume $g_{i}=g$ in this paper for simplicity. For our purposes we also
assume the weak-driving fields specified by
\begin{equation}
|\Omega _{1,2}(t)|\ll g.  \label{e7}
\end{equation}
This assumption neglects the probability of populating state $|\phi
_{3}\rangle $ during the entire process of evolution. Then, at an instant
time $t$ the eigenstates $|\psi _{n}(t)\rangle $ and eigenvalues $\lambda
_{n}(t)$ of $H_{0}(t),$ which obey the equation $H_{0}(t)|\psi
_{n}(t)\rangle =\lambda _{n}(t)|\psi _{n}(t)\rangle ,$ can be derived
analytically. For the former,
\begin{eqnarray}
|\psi _{1}(t)\rangle &=&\left( \Omega _{1}^{2}(t)+\Omega _{2}^{2}(t)+\frac{%
\Omega _{1}^{2}(t)\Omega _{2}^{2}(t)}{g^{2}}\right) ^{-1/2}  \nonumber \\
&&\left( -i\Omega _{2}(t)|\phi _{1}\rangle +\frac{i\Omega _{1}(t)\Omega
_{2}(t)}{g}|\phi _{3}\rangle \right.  \nonumber \\
&& +\Omega _{1}(t)|\phi _{5}\rangle \bigg) ,  \label{e8}
\end{eqnarray}
\begin{eqnarray}
|\psi _{2}(t)\rangle &\simeq &\frac{i\Omega _{1}(t)}{\sqrt{2\left( \Omega
_{1}^{2}(t)+\Omega _{2}^{2}(t)\right) }}|\phi _{1}\rangle -\frac{i}{2}|\phi
_{2}\rangle  \nonumber  \label{e9} \\
&&+\frac{i}{2}|\phi _{4}\rangle +\frac{\Omega _{2}(t)}{\sqrt{2\left( \Omega
_{1}^{2}(t)+\Omega _{2}^{2}(t)\right) }}|\phi _{5}\rangle ,
\end{eqnarray}
\begin{eqnarray}
|\psi _{3}(t)\rangle &\simeq &\frac{i\Omega _{1}(t)}{\sqrt{2\left( \Omega
_{1}^{2}(t)+\Omega _{2}^{2}(t)\right) }}|\phi _{1}\rangle +\frac{i}{2}|\phi
_{2}\rangle  \nonumber  \label{e10} \\
&&-\frac{i}{2}|\phi _{4}\rangle +\frac{\Omega _{2}(t)}{\sqrt{2\left( \Omega
_{1}^{2}(t)+\Omega _{2}^{2}(t)\right) }}|\phi _{5}\rangle ,
\end{eqnarray}
\begin{eqnarray}
|\psi _{4}(t)\rangle &\simeq &\left( \frac{\Omega _{1}^{2}(t)+\Omega
_{2}^{2}(t)}{g^{2}}+8\right) ^{-1/2}  \nonumber  \label{e11} \\
&&\times \left( -i\frac{\Omega _{1}(t)}{g}|\phi _{1}\rangle +i\sqrt{2}|\phi
_{2}\rangle \right.  \nonumber \\
&&\left. -2i|\phi _{3}\rangle +i\sqrt{2}|\phi _{4}\rangle +\frac{\Omega _{2}%
}{g}(t)|\phi _{5}\rangle \right) ,
\end{eqnarray}
and
\begin{eqnarray}
|\psi _{5}(t)\rangle &\simeq &\left( \frac{\Omega _{1}^{2}(t)+\Omega
_{2}^{2}(t)}{g^{2}}+8\right) ^{-1/2}  \nonumber \\
&&\times \left( -i\frac{\Omega _{1}(t)}{g}|\phi _{1}\rangle -i\sqrt{2}|\phi
_{2}\rangle \right.  \nonumber \\
&&\left. -2i|\phi _{3}\rangle -i\sqrt{2}|\phi _{4}\rangle +\frac{\Omega
_{2}(t)}{g}|\phi _{5}\rangle \right) ,  \label{e12}
\end{eqnarray}
and, for the latter, $\lambda _{1}=0,\ \lambda _{2}\simeq -\sqrt{[\Omega
_{1}^{2}(t)+\Omega _{2}^{2}(t)]/2},\ \lambda _{3}\simeq \sqrt{[\Omega
_{1}^{2}(t)+\Omega _{2}^{2}(t)]/2},\ \lambda _{4}\simeq -\sqrt{2}g\ $and $%
\lambda _{5}\simeq \sqrt{2}g,$ respectively. Note that the eigenstate $|\psi
_{1}(t)\rangle $ with zero eigenvalue $\lambda _{1}=0$ is a dark state.

Here, starting from the atomic state $|g\rangle _{1}|f\rangle _{2},$ we are
concerned with two problems. The first problem is how to transfer population
simultaneously in the two atoms: $|g\rangle _{1}\rightarrow |f\rangle _{1}$
and $|f\rangle _{2}\rightarrow |g\rangle _{2}$ or, in other words, how to
drive the two atoms from the initial state $|g\rangle _{1}|f\rangle _{2}$ to
the target state $|f\rangle _{1}|g\rangle _{2}.$ The second problem is how
to create maximum entanglement between the two atoms. As is well known, the
adiabatic passage method does well with these problems. Namely, when the
adiabatic condition \cite{PRA-40-6741-1989} $\left| \left\langle \psi
_{n\neq 1}(t)\right. |\partial _{t}\psi _{1}(t)\rangle \right| \ll \left|
\lambda _{n\neq 1}\right| ,$ with $\partial _{t}\equiv \partial /\partial t,$
is satisfied, state $|\psi _{1}(0)\rangle $ would follow $|\psi
_{1}(t)\rangle $ closely. Then, by judiciously tailoring the classical
fields $\Omega _{1}(t)$ and $\Omega _{2}(t),$ either of the two
above-mentioned problems can be solved successfully. Although the adiabatic
passage method is one-step implementation, it usually takes quite a long
time that is undesirable. If one attempts to quicken the process a little
bit, the adiabatic following condition may be violated and transition to
states other than $|\psi _{1}(t)\rangle $ may happen leading to a wrong
(unintended) final state.

In the next sections we shall consider fast and robust shortcuts to
adiabaticity for the two above problems.

\section{Shortcut for simultaneous population transfer in two atoms}

As the instantaneous eigenstates $\{|\psi _{n}(t)\rangle \},$ Eqs. (\ref{e8}%
) - (\ref{e12}), are not solutions of Schrodinger equation $i\partial
_{t}|\psi _{n}(t)\rangle =H_{0}(t)|\psi _{n}(t)\rangle ,$ there is a finite,
though small, probability that the system starts from state $|\psi
_{n}(0)\rangle $ and ends up in state $|\psi _{m\neq n}(t)\rangle ,$ even
under the adiabatic following condition. To guarantee zero transition
probability for $|\psi _{n}(0)\rangle \rightarrow |\psi _{m\neq n}(t)\rangle
,$ we look for a Hamiltonian $H(t)$ that is related to the original
Hamiltonian $H_{0}(t)$ but drives the eigenstates $\{|\psi _{n}(t)\rangle \}$
exactly, i.e., $i\partial _{t}|\psi _{n}(t)\rangle =H(t)|\psi _{n}(t)\rangle
.$

According to Berry's general transitionless tracking algorithm \cite
{JPA-42-365303-2009}, one can reverse engineer $H(t)$ from $H_{0}(t).$ The
algorithm results in infinitely many such Hamiltonians $H(t)$ which differ
from each other only by phases. Disregarding the effect of phases, the
simplest Hamiltonian $H_{1}(t)$ that exactly drives the set of instantaneous
eigenstates of $H_{0}(t)$ is derived in the form
\begin{equation}
H_{1}(t)=i\sum_{m=1}^{5}|\partial _{t}\psi _{m}(t)\rangle \langle \psi
_{m}(t)|.  \label{e13}
\end{equation}
The addition of $H_{0}(t)$ to $H_{1}(t)$ (i.e., $H(t)=H_{0}(t)+H_{1}(t))$
only affects the phases of the system evolution. Being interested only in
populations, we can exclude $H_{0}(t)$ (i.e., $H(t)=H_{1}(t))$. Putting Eqs.
(\ref{e8}) - (\ref{e12}) into Eq. (\ref{e13}), we obtain, after
differentiating each of the $|\psi _{n}(t)\rangle $ and then summing up all
the five terms, the following expression for $H_{1}(t):$%
\begin{equation}
H_{1}(t)=C(t)\left( |\phi _{1}\rangle \langle \phi _{5}|+|\phi _{5}\rangle
\langle \phi _{1}|\right) ,  \label{e14}
\end{equation}
where
\begin{equation}
C(t)=\frac{\Omega _{1}(t)\partial _{t}\Omega _{2}(t)-\Omega _{2}(t)\partial
_{t}\Omega _{1}(t)}{\Omega _{1}^{2}(t)+\Omega _{2}^{2}(t)+\Omega
_{1}^{2}(t)\Omega _{2}^{2}(t)/g^{2}}.  \label{e15}
\end{equation}

We remark, however, that for two $\Lambda $-type atoms in real experiment,
the CDD Hamiltonian $H_{1}(t)$ in Eq. (\ref{e14}) does not exist. Therefore,
we shall find an alternative physically feasible Hamiltonian which is
equivalent to $H_{1}(t).$ Generally, the physical realization of such
Hamiltonians is case-dependent. For example, in Chen's scheme \cite
{PRL-105-123003-2010} for nonadiabatic speeding up the population transfer
in two- and three-level systems of a single atom, an auxiliary laser or
microwave interactions are involved to directly drive two internal levels of
the atom. Here we take into account an auxiliary excited level $|e\rangle
_{i},$ additional classical driving fields $\tilde{\Omega}_{i}(t)$ and an
extra cavity field mode to realize an equivalent-to-the-CDD Hamiltonian that
indirectly drives the two interested states $|\phi _{1}\rangle $ and $|\phi
_{5}\rangle ,$ as shown by dashed bars and dashed arrows in Fig. 1. The
transition $|f\rangle _{i}\leftrightarrow |e\rangle _{i}$ is dispersively
coupled to the auxiliary cavity mode with a real coupling constant $\tilde{g}%
_{i}$ and a detuning $\Delta _{2}$, while $|g\rangle _{i}$ is nonresonantly
coupled to $|e\rangle _{i}$ by a laser field with a Rabi frequency $\tilde{%
\Omega}_{i}(t)$ and a detuning $\Delta _{1}$. Under the RWA, the auxiliary
interaction Hamiltonian is $(\hbar =1)$
\begin{equation}
\tilde{H}(t)=\sum_{i=1}^{2}\left( \tilde{\Omega}_{i}(t)e^{i\Delta
_{1}t}|g\rangle _{i}\langle e|+\tilde{g}_{i}e^{i\Delta _{2}t}b^{\dag
}|f\rangle _{i}\langle e|+H.c.\right) ,  \label{e16}
\end{equation}
where $b^{\dag }$ is the creation operation for the auxiliary cavity mode.
We assume $\tilde{g}_{i}=g$ and $\tilde{\Omega}_{i}(t)=\tilde{\Omega}(t)$ in
this paper for simplicity.

Let the system be initially in state $|\phi _{1}\rangle .$ In the large
detuning regime $\Delta _{1},\Delta _{2}\gg \tilde{\Omega}(t),g$ and $%
|\delta |\equiv |\Delta _{1}-\Delta _{2}|\gg \eta (t)\equiv \frac{1}{2}(%
\frac{1}{\Delta _{1}}+\frac{1}{\Delta _{2}})g\tilde{\Omega}(t),$ the atoms
can mutually exchange energy in such a way that the level $|e\rangle $ and
the auxiliary cavity field mode $b$ are only virtually excited \cite
{APL-94-154101-2009,EPJD-61-737-2011}. Then $\tilde{H}(t)$ can effectively
be described by the Hamiltonian
\begin{equation}
\tilde{H}_{eff}(t)=\frac{\eta ^{2}(t)}{\delta }%
(S_{1}^{+}S_{2}^{-}+S_{2}^{+}S_{1}^{-}),  \label{e17}
\end{equation}
where $S_{j}^{+}=|f\rangle _{j}\langle g|\ $and $S_{j}^{-}=|g\rangle
_{j}\langle f|,$ with $j=1,2.$ The effective Hamiltonian (\ref{e17}) is
equivalent with the CDD Hamiltonian $H_{1}(t)$ in Eq. (\ref{e14}) when
\begin{equation}
\frac{\eta ^{2}(t)}{\delta }=C(t).  \label{e18}
\end{equation}
Hence, the Rabi frequency of the auxiliary laser field that generates a
Hamiltonian equivalent to the CDD Hamiltonian can be determined from the
original frequencies $\Omega _{1}(t)$ and $\Omega _{2}(t)$ as
\begin{equation}
\tilde{\Omega}(t)=\frac{2\Delta _{1}\Delta _{2}}{\Delta _{1}+\Delta _{2}}%
\sqrt{\frac{\left( \Omega _{1}(t)\partial _{t}\Omega _{2}(t)-\Omega
_{2}(t)\partial _{t}\Omega _{1}(t)\right) \delta }{\left( \Omega
_{1}^{2}(t)+\Omega _{2}^{2}(t)\right) g^{2}+\Omega _{1}^{2}(t)\Omega
_{2}^{2}(t)}}.  \label{e19}
\end{equation}
We remark that, compared to the scheme in Ref.
\cite{APL-94-154101-2009}, we will prepare the maximal entanglement
in the ground states of two atoms, which is robust against the
atomic spontaneous emission. Moreover, different from the scheme in
Ref. \cite{APL-94-154101-2009}, the auxiliary Hamiltonian
(\ref{e17}) can serve as a supplement or function independently for
our time-dependent CDD Hamiltonian. When the auxiliary Hamiltonian
functions independently, our system reduces to the interaction
between two simple three-level atomic systems and a field mode which
could be realized with the current cavity QED technology. Thus,
based on the CDD Hamiltonian (\ref {e17}), our proposals for fast
population transfer and creation of two-atom maximal entanglement
are not sensitive to fluctuations of the experiment parameters.
Especially, there is no need to precisely control the operation time
in our present scheme.

\begin{figure}[tbp]
\centering{\includegraphics[width=0.8\columnwidth]{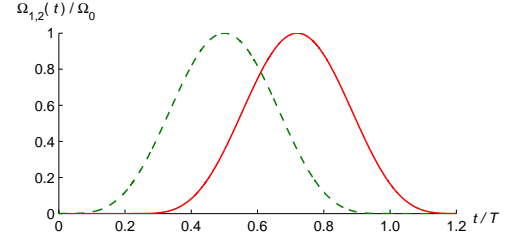}}
\caption{The dependence on $t/T$ of $\Omega _{1}(t)/\Omega _{0}$ (solid red
line) and $\Omega _{2}(t)/\Omega _{0}$ (dashed green line), where $\Omega
_{1}(t)$ and $\Omega _{2}(t)$ are defined by Eqs. (\ref{e20}) and (\ref{e21}%
), for the population transfer, with$\ \tau =0.22T.$}
\label{Fig.2}
\end{figure}

\begin{figure}[tbp]
\centering
\subfigure[]{\label{Fig.sub.a}\includegraphics[width=0.8\columnwidth]{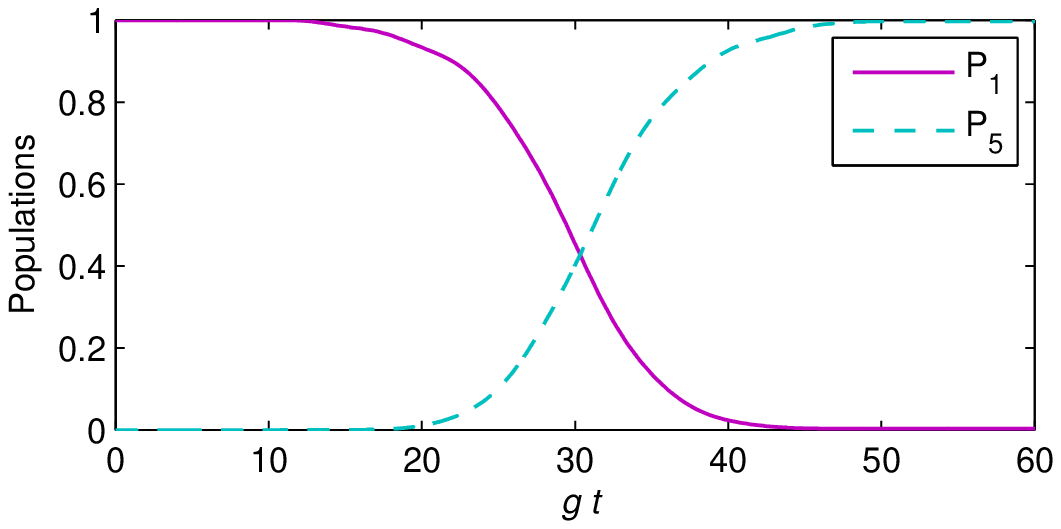}%
} \subfigure[]{\label{Fig.sub.b}\includegraphics[width=0.8\columnwidth]{%
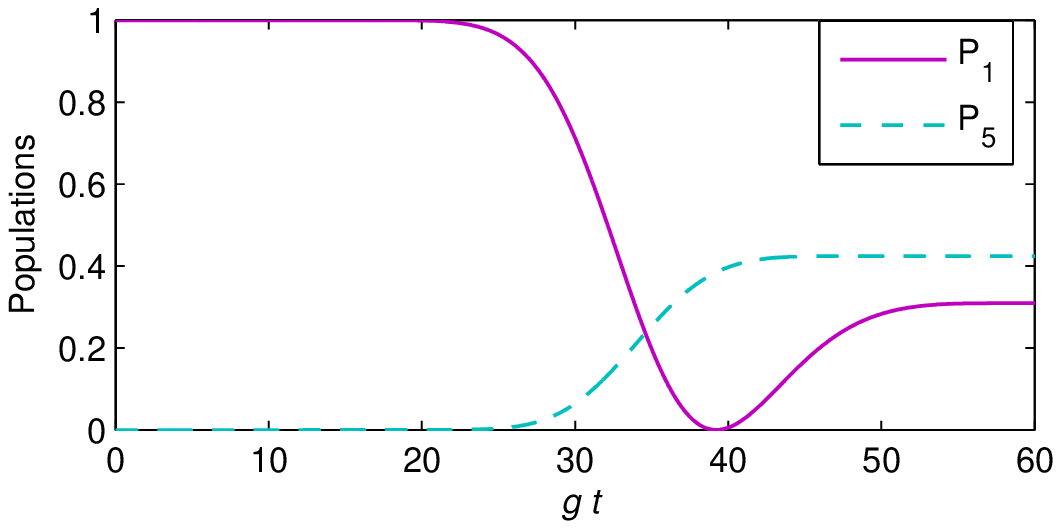}}
\caption{The dependence on $gt$ of the populations $P_{1}(t)$ and $P_{5}(t)$
governed by (a) the auxiliary interaction Hamiltonian $\tilde{H}(t)$ with $%
\Delta _{1}=6g,$ $\Delta _{2}=7g$ and (b) the original Hamiltonian $%
H_{0}(t). $ In both (a) and (b) the Rabi frequencies $\Omega _{1}(t)$ and $%
\Omega _{2}(t)$ are defined by Eqs. (\ref{e20}) and (\ref{e21}) with $\Omega
_{0}=0.2g,\ T=50/g$ and $\tau =0.22T.$ }
\end{figure}

\begin{figure}[tbp]
\centering{\includegraphics[width=0.8\columnwidth]{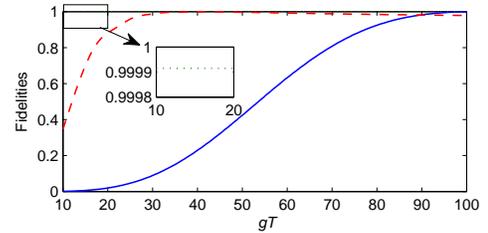}}
\caption{The dependence on $gT$ of the fidelities of the evolved state with
respect to the target state using adiabatic passage governed by the
Hamiltonian $H_{0}(t)$ (solid blue line), the dynamics governed by the CDD
Hamiltonian $H_{1}(t)$ (dotted green line), and by the auxiliary interaction
Hamiltonian $\tilde{H}(t)$ (dashed red line) with $\Delta _{1}=6g,$ $\Delta
_{2}=7g$ for the case of population transfer. The Rabi frequencies $\Omega
_{1}(t)$ and $\Omega _{2}(t)$ are defined by Eqs. (\ref{e20}) and (\ref{e21}%
) with $\Omega _{0}=0.2g$ and $\tau =0.22T.$}
\label{Fig.4}
\end{figure}

We now demonstrate that the simultaneous population transfer $|g\rangle
_{1}|f\rangle _{2}\rightarrow |f\rangle _{1}|g\rangle _{2}$ in the two atoms
governed by\textbf{\ $\tilde{H}(t)$} is much speeded up as compared to the
adiabatic passage governed by $H_{0}(t).$ Let the Rabi frequencies $\Omega
_{1}(t)$ and $\Omega _{2}(t)$ in the original Hamiltonian $H_{0}(t)$ depend
on time as
\begin{equation}
\Omega _{1}(t)=\left\{
\begin{tabular}{ll}
$\Omega _{0}\sin ^{4}[\pi (t-\tau )/T]$ & for $\tau \leq t\leq T+\tau $ \\
$0$ & otherwise
\end{tabular}
\right.  \label{e20}
\end{equation}
and
\begin{equation}
\Omega _{2}(t)=\left\{
\begin{tabular}{ll}
$\Omega _{0}\sin ^{4}(\pi t/T)$ & for $0\leq t\leq T$ \\
$0$ & otherwise
\end{tabular}
\right. ,  \label{e21}
\end{equation}
with $\Omega _{0}$ being the pulse amplitude, $\tau $ being the time delay
and $T$ being the operation duration. Fig. 2 shows $\Omega _{1}(t)/\Omega
_{0}$ and $\Omega _{2}(t)/\Omega _{0}$ as functions of $t/T$ for a fixed
value of the time delay chosen for the best adiabatic passage process. With $%
\Omega _{1}(t)$ and $\Omega _{2}(t)$ defined in (\ref{e20}) and (\ref{e21}),
we contrast the performances of population transfer from the initial state $%
|\phi _{1}\rangle $ to the target state $|\phi _{5}\rangle $ based on the
adiabatic passage method governed by $H_{0}(t)$ and on the evolution
governed by our auxiliary interaction Hamiltonian $\tilde{H}(t)$ in Fig. 3,
where populations $P_{k}(t)=\langle \phi _{k}|\rho (t)|\phi _{k}\rangle ,$
with $\rho (t)$ being the density matrix associated with the governing
Hamiltonian, are plotted versus $gt$ for a given operation duration. The
result obviously reveals that near-perfect population transfer by\textbf{\ $%
\tilde{H}(t)$} can be achieved even in a short evolution time (see Fig. 3a)
for which the adiabatic passage method breaks down (see Fig. 3b). This means
that \textbf{$\tilde{H}(t)$ }indeed provides a shortcut to the adiabatic
passage.

In Fig. 4, we plot the fidelity $F(T+\tau)=\left( \text{Tr}\sqrt{\rho
_{f}^{1/2}\rho (T+\tau)\rho _{f}^{1/2}}\right) ^{2},$ with $\rho _{f}$ being
the density matrix of the ideal final state and $\rho (T+\tau)$ being that
of the evolved state at the end of the pulse operation, as a function of the
operation time $T,$ by using the original Hamiltonian $H_{0}(t)$, the CDD
Hamiltonian $H_{1}(t)$ and the auxiliary interaction Hamiltonian $\tilde{H}%
(t).$ While $H_{1}(t)$ formally yields a near-perfect population transfer
within an arbitrarily short time, \textbf{$\tilde{H}(t)$ }needs a finite
operation time to complete the population transfer. This is because $\tilde{H%
}_{eff}(t)$ is valid only under the large-detuning condition, that requires
longer evolution time to achieve a full population transfer than that
required in the resonant-driving CDD Hamiltonian $H_{1}(t)$. For example,
here the fidelity of the evolved state would be higher than $98\%$ when $%
T\geq 30/g.$ As for the adiabatic passage evolution via $H_{0}(t),$ it
requires much more time to finish the population transfer: it is about three
times longer than that required by\textbf{\ $\tilde{H}(t)$}.

\begin{figure}[tbp]
\centering{\includegraphics[width=0.8\columnwidth]{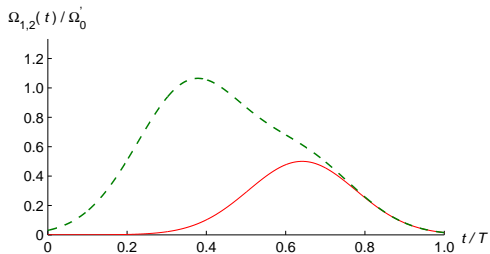}}
\caption{$\Omega _{1}(t)/\Omega _{0}^{\prime }$ (solid red line) and $\Omega
_{2}(t)/\Omega _{0}^{\prime }$ (dashed green line) versus $t/T$ for the
creation of maximal entanglement between two atoms, where $\Omega _{1}(t)$
and $\Omega _{2}(t)$ are defined by Eqs.(\ref{e22}) and (\ref{e23}) with $%
\theta =17/120$ and $w=23/120.$}
\label{Fig.5}
\end{figure}

\section{Shortcut for maximal entanglement creation between two atoms}

\begin{figure}[tbp]
\centering\subfigure[]{\label{Fig.sub.a}\includegraphics[width=0.8%
\columnwidth]{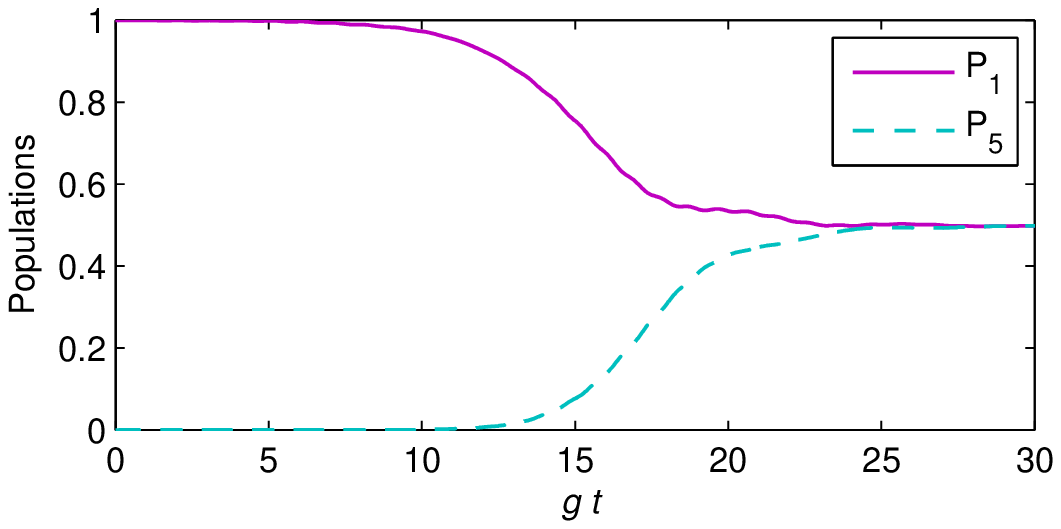}} \subfigure[]{\label{Fig.sub.b}%
\includegraphics[width=0.8\columnwidth]{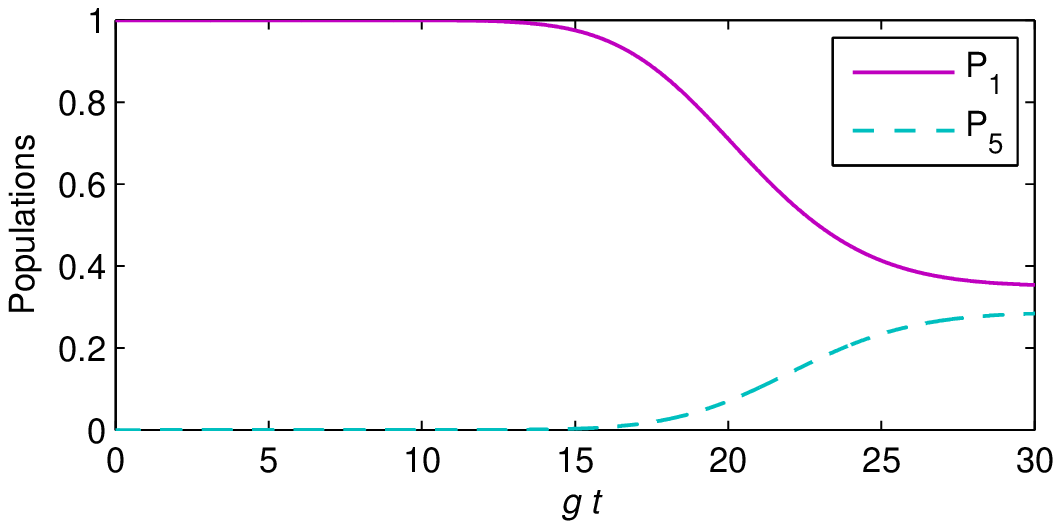}}
\caption{The dependence on $gt$ of the populations $P_{1}(t)$ and $P_{5}(t)$
governed by (a) the auxiliary interaction Hamiltonian $\tilde{H}(t)$ with $%
\Delta _{1}=6g,$ $\Delta _{2}=7g$ and (b) the original Hamiltonian $%
H_{0}(t). $ In both (a) and (b) the Rabi frequencies $\Omega _{1}(t)$ and $%
\Omega _{2}(t)$ are defined by Eqs. (\ref{e22}) and (\ref{e23}) with $\Omega
_{0}^{\prime }=0.3g,$ $\theta =17/120,$ $w=23/120$ and $T=30/g.$}
\label{Fig.6}
\end{figure}

\begin{figure}[tbp]
\centering{\includegraphics[width=0.8\columnwidth]{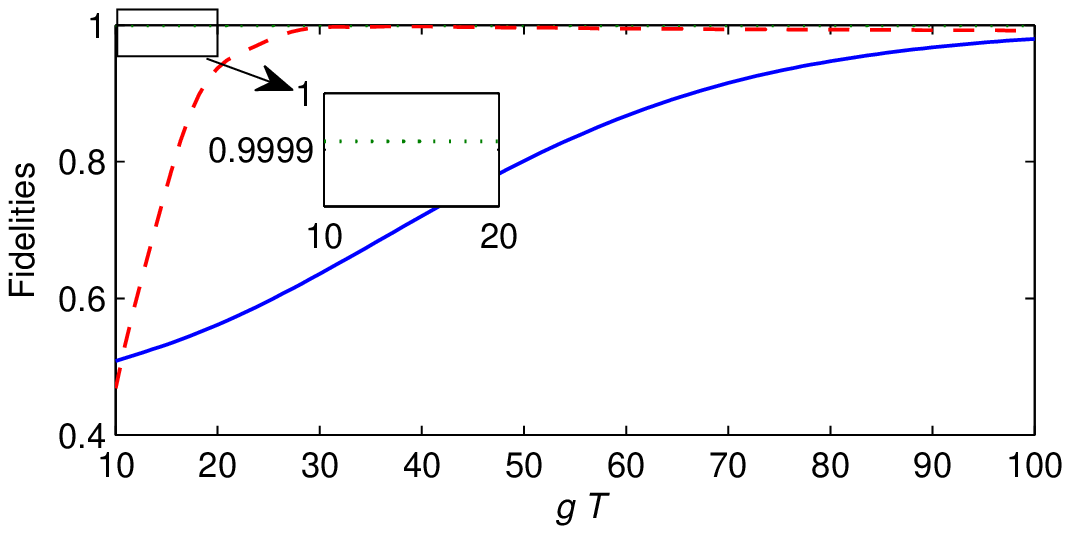}}
\caption{The dependence on $gT$ of the fidelities of the evolved state with
respect to the target state using adiabatic passage governed by the
Hamiltonian $H_{0}(t)$ (solid blue line), the dynamics governed by the CDD
Hamiltonian $H_{1}(t)$ (dotted green line), and by the auxiliary interaction
Hamiltonian $\tilde{H}(t)$ (dashed red line) with $\Delta _{1}=6g,$ $\Delta
_{2}=7g$ for the case of maximal entanglement creation. The Rabi frequencies
$\Omega _{1}(t)$ and $\Omega _{2}(t)$ are defined by Eqs. (\ref{e22}) and (%
\ref{e23}) with $\Omega _{0}^{\prime }=0.3g,\theta =17/120$ and $w=23/120.$}
\label{Fig.7}
\end{figure}
\label{tital}

To our knowledge, speeding up adiabatic passage for creation of maximal
entanglement between two atoms has not been dealt with yet. Under the
weak-driving condition in Eq. \ref{e7}, one should tailor the two Rabi
frequencies $\Omega _{1}(t)$ and $\Omega _{2}(t)$ so that at the beginning
of the operation $\Omega _{1}(t)/\Omega _{2}(t)\rightarrow 0$ but at the end
$\Omega _{1}(t)/\Omega _{2}(t)\rightarrow 1.$ Such requirements can be met
by
\begin{equation}
\Omega _{1}(t)=\frac{1}{2}\Omega _{0}^{^{\prime }}\exp \frac{-[t-(\theta
+1/2)T]^{2}}{w^{2}T^{2}}  \label{e22}
\end{equation}
and
\begin{eqnarray}
\Omega _{2}(t) &=&\Omega _{0}^{^{\prime }}\left[ \exp \frac{-[t+(\theta
-1/2)T]^{2}}{w^{2}T^{2}}\right.  \nonumber \\
&&\left. +\frac{1}{2}\exp \frac{-[t-(\theta +1/2)T]^{2}}{w^{2}T^{2}}\right]
\label{e23}
\end{eqnarray}
with $\Omega _{0}^{\prime }$ being the pulse amplitude, $T$ being the
operation duration, while $\theta $ and $w$ being some parameters to be
chosen for the best performance of the adiabatic passage process. The
time-dependences of $\Omega _{1}(t)/\Omega _{0}^{\prime }$ and $\Omega
_{2}(t)/\Omega _{0}^{\prime }$ are shown in Fig. 5 versus $t/T$ for fixed
values of $\theta $ and $w.$ With such tailored $\Omega _{1}(t)$ and $\Omega
_{2}(t),$ we plot in Fig. 6 the evolution of populations $P_{1,5}(t),$ based
on the dynamics governed by the interaction Hamiltonian $\tilde{H}(t),$ Fig.
6a, and on the adiabatic passage governed by the original Hamiltonian $%
H_{0}(t),$ Fig. 6b. From Fig. 6a, we see that at $t\simeq 30/g$ the
interaction Hamiltonian $\tilde{H}(t)$ already yields $P_{1}(t)\approx
P_{5}(t)=1/2.$ That is, the system state at $t\simeq 30/g$\ becomes\textbf{\
$\left( -i|\phi _{1}\rangle +|\phi _{5}\rangle \right) /\sqrt{2},$}
signifying creation of the atoms' maximal entangled state
\begin{equation}
|\Phi \rangle _{12}=\frac{1}{\sqrt{2}}(-i|g\rangle _{1}|f\rangle
_{2}+|f\rangle _{1}|g\rangle _{2}).  \label{e24}
\end{equation}
From Fig. 6b, however, governed by the original Hamiltonian $H_{0}(t),$ $%
P_{1}(t\simeq 30/g)$ is still quite bigger than $P_{5}(t\simeq 30/g).$

In Fig. 7 we also plot the fidelities of the evolved states governed by $%
\tilde{H}(t)$ and $H_{0}(t),$ with respect to the maximally entangled state $%
|\Phi \rangle _{12},$ as functions of the operation duration $T.$ Clearly
from Fig. 7, to obtain the maximum entanglement between the two atoms, a
much longer operation time is required by adiabatic passage method than that
by\textbf{\ $\tilde{H}(t).$} Thus, the results in Figs. 6 and 7 confirm that%
\textbf{\ $\tilde{H}(t)$} indeed accelerates the creation of maximally
entangled state for the two atoms as compared to the original Hamiltonian $%
H_{0}(t).$ Or, in other words, the dynamics governed by $\tilde{H}(t)$ is a
nonadiabatic shortcut to adiabaticity governed by $H_{0}(t)$ for creation of
maximal entanglement between two atoms within a cavity.

\section{Robustness of the shortcut schemes}

Not only speed but also robustness against possible mechanisms of
decoherence is important for a scheme to be applicable in quantum
information processing and quantum computing. In the problems of our concern
here decoherences may originate from the atomic spontaneous emission and the
cavity decay. To examine robustness of our shortcut schemes described in the
previous sections against such decoherence mechanisms we numerically solve
the master equation for the whole system's density matrix $\rho (t),$ which
has the form
\begin{eqnarray}
\partial _{t}\rho (t) &=&-i[H_{0}(t)+\tilde{H}(t),\rho (t)]  \nonumber \\
&&-\frac{\kappa _{a}}{2}(a^{\dag }a\rho (t)-2a\rho (t)a^{\dag }+\rho
(t)a^{\dag }a)  \nonumber \\
&&-\frac{\kappa _{b}}{2}(b^{\dag }b\rho (t)-2b\rho (t)b^{\dag }+\rho
(t)b^{\dag }b)  \nonumber \\
&&-\sum_{k={1}}^{2}\sum_{m=g,f}\sum_{n=s,e}\frac{\Gamma _{nm}^{k}}{2}%
[S_{mn}^{k+}S_{mn}^{k}\rho (t)  \nonumber \\
&&-2S_{mn}^{k}\rho (t)S_{mn}^{k+}+\rho (t)S_{mn}^{k+}S_{mn}^{k}],
\label{e25}
\end{eqnarray}
where $\kappa _{a}$ $(\kappa _{b})$ is the photon leakage rate of
the cavity mode $a$ $(b),$ $\Gamma _{nm}^{k}$ is the $k$th atom's
spontaneous emission rate from the excited state $|n\rangle _{k}$ to
the ground state $|m\rangle _{k}$ and $S_{mn}^{k}=|m\rangle
_{k}\langle n|=S_{nm}^{k+}.$ For simplicity, we assume $\Gamma
_{nm}^{k}=\Gamma /2$ and $\kappa _{a}=\kappa _{b}=\kappa $ in
numerically solving the master equation (\ref{e25}) with the initial
condition $\rho (0)=\left| \phi _{1}\right\rangle \left\langle \phi
_{1}\right| .$ In Fig. 8, we display the dependence on the ratios
$\Gamma /g$ and $\kappa /g$ of the fidelities of the evolved state
at the end of the operation time for the population transfer (Fig.
8a) and the creation of maximal entanglement (Fig. 8b). The physical
configuration that we consider in the present scheme employs
electric dipole transitions on the $D_{1}$ line of a single
$^{87}Rb$ atom \cite{Rb-atom}. In real experiment, one can couple
this atom simultaneously to two optical cavity modes and two laser
fields. Two stable hyperfine ground states are the ($F=1,m=-1$)
level and the ($F=2,m=-2$) level of the $5^{2}S_{1/2}$ state, while
two metastable hyperfine excited states are the ($F^{^{\prime
}}=1,m=-1$) level and the ($F^{^{\prime }}=2,m=-2$) level of
$5^{2}P_{1/2}.$ The related cavity-QED parameters could be
achievable with, for example, microtoroidal whispering-gallery-mode
resonators \cite{PRA-71-013817-2005}.
In current experiments, the parameters $g=2.5GHz,\ \kappa =10MHz$ and $%
\Gamma =10MHz$ have been reported in Ref. \cite
{PRA-71-013817-2005,NP-2-849-2006}. For such parameters, we can see from
Fig. 8 that fidelities higher than $98\%$can be achieved in both the
shortcut schemes. Therefore, the schemes are robust and might be promising
within the current technology.

\begin{figure}[tbp]
\centering\subfigure[]{\label{Fig.sub.a}\includegraphics[width=0.8%
\columnwidth]{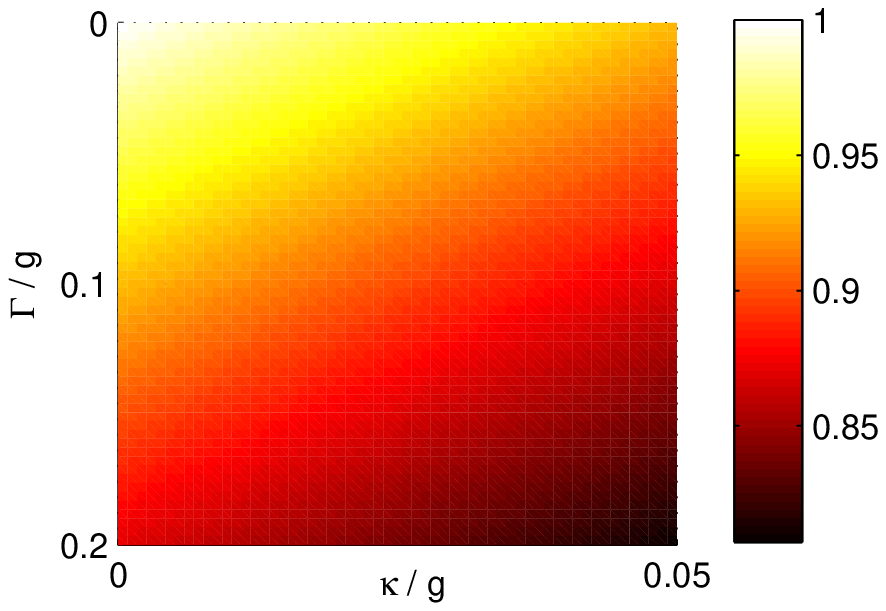}} \subfigure[]{\label{Fig.sub.b}%
\includegraphics[width=0.8\columnwidth]{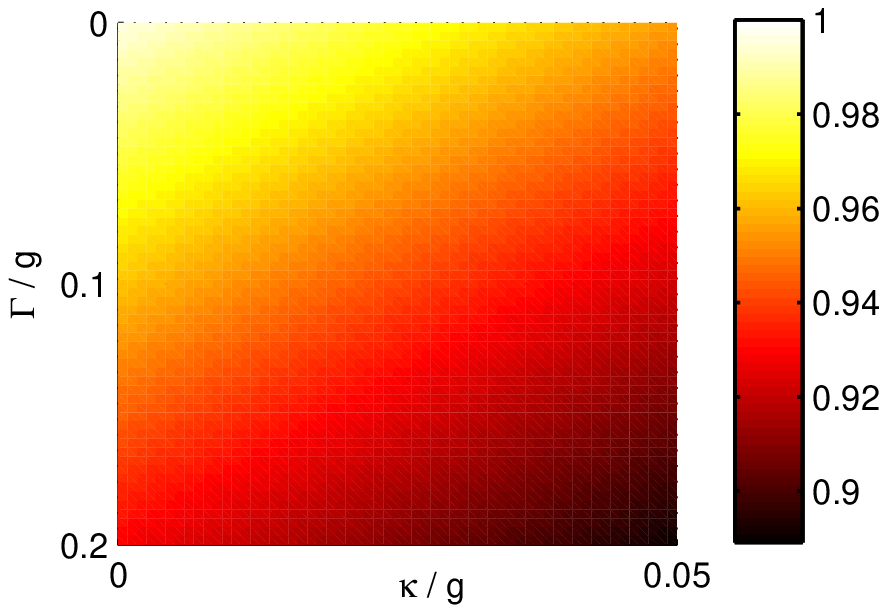}}
\caption{The dependences on $\Gamma /g$ and $\kappa /g$ of the fidelities of
the evolved state at the end of the operation time with respect to the
target state that are obtained by numerically solving the master equation (%
\ref{e15}) for (a) the population transfer with $\Omega _{1}(t)$ and $\Omega
_{2}(t)$ defined by Eqs. (\ref{e20}) and (\ref{e21}) with $\Omega _{0}=0.2g$%
, $T=50/g$ and $\tau =0.22T$ and (b) the maximal entanglement creation with $%
\Omega _{1}(t)$ and $\Omega _{2}(t)$ defined by Eqs. (\ref{e22}) and (\ref
{e23}) with $\Omega_{0}^{\prime }=0.3g,$ $\theta =17/120$, $w=23/120$ and $%
T=30/g$.}
\label{Fig.8}
\end{figure}

\section{Conclusion}

In conclusion, we propose nonadiabatic shortcut schemes for the population
transfer and creation of maximal entanglement between two atoms based on the
CQED that perform much faster than those based on the adiabatic passage
method. Using the transitionless driving approach, we analytically derive a
CDD Hamiltonian, the shortcut performance of which is numerically
demonstrated to be equivalent to our auxiliary interaction Hamiltonian under
large detuning regime. The speed in both the population transfer and the
maximal entanglement creation between two atoms can be improved about three
times those based on the conventional adiabatic passage. Also, the present
schemes are shown robust against the decoherences caused by the atomic
spontaneous emission and cavity decay.

\section{Acknowledgments}

M. L. and Y. X. were supported by the National Natural Science
Foundation of China under grant no. 11105030 and no. 11374054, the
Major State Basic Research Development Program of China under Grant
No. 2012CB921601. S. J. was supported by the National Natural
Science Foundation of China under grant no.11205037.  N. B. A. was
funded by Vietnam National Foundation for Science and Technology
Development (NAFOSTED).

\end{document}